\documentclass[a4paper,11pt]{article}
\pdfoutput=1 

\usepackage{jcappub} 

\usepackage[T1]{fontenc} 

\DeclareUnicodeCharacter{2212}{-}
\usepackage{comment}

\makeatother   
\newcommand{\GeV}{{\rm \,GeV}}
\newcommand{\TeV}{{\rm \,TeV}}
\newcommand{\MeV}{{\rm \,MeV}}

\newcommand{\cm}{{\rm \,cm}}

\newcommand{\g}{{\rm \,g}}
\newcommand{\K}{{\rm \,K}}
\newcommand{\yr}{{\rm \,yr}}

\newcommand{\Gyr}{{\rm \,Gyr}}

\newcommand{\Msun}{M_\odot}
\newcommand{\Mstar}{M_\star}
\newcommand{\Rstar}{R_\star}
\newcommand{\vstar}{v_\star}

\newcommand{\Tstar}{T_\star}

\newcommand{\Teff}{T_{\rm eff}}

\newcommand{\vesc}{v_\text{esc}}
\newcommand{\umax}{u_\chi^\text{max}}

\newcommand{\fMB}{f_{\rm MB}}

\begin{document}

\hfill FERMILAB-PUB-25-0343-T,  LAPTH-020/25

\title{Can local White Dwarfs constrain Dark Matter interactions?}

\author[a]{Pooja Bhattacharjee,}
\author[b, c]{Sandra Robles,}
\author[d, e]{Stephan~A.~Meighen-Berger}
\author[f]{and Francesca Calore}

\affiliation[a]{Center for Astrophysics and Cosmology, University of Nova Gorica, Vipavska 13, SI-5000 Nova Gorica, Slovenia}
\affiliation[b]{Particle Theory Department, Theory Division, Fermi National Accelerator Laboratory, Batavia, Illinois 60510, USA}
\affiliation[c]{Kavli Institute for Cosmological Physics, University of Chicago, Chicago, Illinois 60637, USA}
\affiliation[d]{School of Physics, The University of Melbourne, Victoria 3010, Australia}
\affiliation[e]{Center for Cosmology and AstroParticle Physics (CCAPP), Ohio State University, Columbus, OH 43210, USA}
\affiliation[f]{Laboratoire d'Annecy-le-Vieux de Physique Théorique, CNRS, USMB, F-74940 Annecy, France}

\emailAdd{pooja.bhattacharjee@ung.si}
\emailAdd{srobles@fnal.gov}
\emailAdd{stephan.meighenberger@unimelb.edu.au}
\emailAdd{calore@lapth.cnrs.fr}

\abstract{
We investigate whether nearby white dwarfs (WDs) can constrain dark matter (DM) interactions with ordinary matter. As experimental sensitivity improves,  driven by the Gaia mission, the sample volume of nearby WDs has been increasing over recent years. We carefully select a sample of ten cold, isolated, non-magnetic WDs within 13~pc of the Sun. We model their carbon-oxygen interior using a finite temperature relativistic equation of state and model atmospheres to infer their core temperatures. This enables us to perform a thorough estimation of the DM capture rate and evaporation mass using actual astrophysical observations. Given the low local DM density, we focus on DM that annihilates into long-lived mediators, which escape the WD and later decay into photons. While \textit{Fermi}-LAT data shows no significant gamma-ray excess, future telescopes, CTA North \& South, LHAASO, SWGO, could probe DM-nucleon cross sections down to $\sim 10^{-41}~\text{cm}^2$ for DM masses above the TeV scale. Our results are competitive with current direct detection bounds (e.g., LZ) in the multi-TeV regime. This work underscores the importance of systematic WD studies in the broader landscape of DM detection and demonstrates the synergy between astrophysical and terrestrial searches in exploring DM interactions.

\noindent }

\maketitle

\section{Introduction}
\label{intro}

\noindent Investigating the potential signatures of dark matter (DM) accumulation in compact stellar objects provides a promising avenue for detecting its presence and understanding its properties. Stellar remnants can accrete considerably more DM than other celestial bodies due to their greater gravitational potential. 
Among them, white dwarfs (WDs) are the most abundant in the Galaxy and their physics is better understood than that of neutron stars. Recently, the Gaia mission~\cite{GentileFusillo:2021, GaiaDR2:2018, GaiaDR3:2020} has yielded a large, well-characterized sample of WD candidates, with precise parallax and photometric measurements, especially in the Solar neighborhood. This has motivated spectroscopy follow-ups, see e.g. refs.~\cite{Kleinman:2013, Kepler:2015, Kepler:2019, OBrien:2024}, which provide the necessary means to infer WD properties such as surface temperature, mass, radius and age.

\noindent WDs offer several compelling advantages as detectors of DM interactions. Their strong gravitational potential allows them to capture DM more efficiently than the Sun, potentially enhancing annihilation signals. In addition, the core of old WDs is significantly colder than that of the Sun, which would allow them to retain captured DM of mass below the GeV scale~\cite{Garani:2021feo, Bell:2021fye}, while DM capture in the Sun is sensitive to DM of mass above $\sim4\GeV$~\cite{Griest:1986yu, Gould:1987ir, Hooper:2008cf, Busoni:2013kaa}. Furthermore, WDs in the Solar neighborhood are both abundant and relatively nearby, making them promising observational targets.

\noindent Various astrophysical consequences of DM accumulation in WDs have been explored in the literature. For non-annihilating DM, proposed effects include triggering supernovae and inducing black hole formation~\cite{Graham:2015apa, Bramante:2015cua, Graham:2018efk, Acevedo:2019gre, Janish:2019nkk, Steigerwald:2022pjo}. In the case of self-annihilating DM, captured particles may contribute to the WD luminosity~\cite{Bertone:2007ae, McCullough:2010ai, Hooper:2010es, Amaro-Seoane:2015uny, Bramante:2017xlb, Dasgupta:2019juq, Panotopoulos:2020kuo, Curtin:2020tkm, Bell:2021fye}. However, for this effect to be observationally significant, the WD must reside in a high-DM-density environment, such as the Galactic Center~\cite{Bell:2021fye}. WDs in globular clusters have also been proposed as potential probes~\cite{Bertone:2007ae, McCullough:2010ai}, provided they not only harbor DM, but also their DM density is ${\cal O}(100\GeV/\cm^3)$. 
However, so far, there is no conclusive evidence of significant DM presence in these stellar systems~\cite{Ibata:2012eq, Carlberg:2021, Reynoso-Cordova:2022ojo, Garani:2023esk}, preventing any meaningful constraints from DM-induced heating in such environments.

\noindent The accumulation of DM in stars proceeds in the following way. As DM particles traverse a star, they can scatter off stellar constituents, losing kinetic energy in the process and becoming gravitationally bound to the star \cite{Gould:1987ir, Gould:1987ju, Goldman:1989nd, Jungman:1995df, Kouvaris:2007ay, Bertone:2007ae, Kouvaris:2010vv,deLavallaz:2010wp, Busoni:2013kaa, Garani:2017jcj, Busoni:2017mhe}. Captured DM continues to scatter with the star constituents until it reaches thermal equilibrium in the star's core. Note that low mass captured DM can gain energy in this process,  and escape the star, thereby reducing the DM population in the core. 
Over time, a significant DM population can accumulate in the core, which in turn may self-annihilate. The byproducts of this annihilation, such as neutrinos~\cite{Super-Kamiokande:2004pou, Tanaka:2011uf, Super-Kamiokande:2015xms, ANTARES:2016xuh, IceCube:2016dgk, Bell:2021esh, IceCube:2021xzo, Bhattacharjee:2023qfi, Bhattacharjee:2024pis} and photons (depending on the DM model)~\cite{Batell:2009zp, Schuster:2009au, Bell:2011sn, Feng:2016ijc, Leane:2017vag, Cermeno:2018qgu, Bell:2021pyy, Bhattacharjee:2022lts, Bhattacharjee:2024pis, Gupta:2025jte}, can escape the star. The potential signal is expected to be strong if an equilibrium between DM accumulation and annihilation is established.

\noindent This paper explores the potential of local WDs as probes of DM interactions with ordinary matter. 
Rather than focusing on an increase in WD luminosity due to DM capture and further self-annihilation, which would not be significant due to the low DM local density and WD internal temperature ${\cal O}(10^6\K)$, 
we investigate a scenario in which captured DM particles annihilate into long-lived mediators~\cite{Rothstein:2009pm, Elor:2015bho, Okawa:2016wrr, Niblaeus:2019gjk}. These mediators escape the WD before decaying into detectable particles namely gamma-ray photons. This setup opens a unique observational window to constrain the DM–nucleon scattering cross section using space- or ground-based telescopes. 
Thus, we aim to identify gamma-ray signatures from nearby WDs and use them to place constraints on the DM–nucleon scattering cross sections in scenarios involving long-lived mediators. We map out the viable parameter space for this mechanism to occur and highlight how it complements existing DM direct searches.

\noindent This paper is organized as follows. In section~\ref{sec:WDs} we briefly outline the WD composition and physics relevant to this work, as well as present our local observational sample. In section~\ref{sec:capture}, we review the calculation of capture   
and annihilation rates as well as the calculation of the photon fluxes from DM annihilation. Our results are presented in section~\ref{sec:gamma_intro} and our conclusions are given in section~\ref{sec:conclusions}.

\section{White Dwarfs}
\label{sec:WDs}
\noindent While the majority of massive stars inevitably undergo supernovas, stars with low or medium masses ($\Mstar\lesssim~8−10 \Msun$), the most abundant in the Universe, deplete their fuel and conclude their life cycle as WDs. The core remnant of a WD is sustained against gravitational collapse through electron degeneracy pressure. WDs constitute the majority of stars in the Galaxy, more than 90\%. Born at high temperatures ${\cal O} (10^8\K)$, WDs cool over billions of years, and the observation of the coldest WDs provides insights into the Galaxy's star formation history.

\subsection{Internal structure}
 
\noindent The internal structure of a WD can be broadly divided into a non-degenerate envelope and a degenerate core. The envelope,  enclosing the WD core, comprises less than 1\% of the total WD mass~\cite{Fontaine:2001}. These outer layers form an atmosphere rich in lighter elements, such as hydrogen and helium, the composition of which evolves as the WD cools. The atmosphere is non-degenerate and highly opaque to radiation and acts as a heat blanketing envelope. 

\noindent Once the core has crystallized, a structure resembling a Coulomb lattice of ions surrounded by degenerate electrons is established, indicating an isothermal and thermally conductive core~\cite{Fontaine:2001, Winget:2008iu, Salaris:2009, Althaus:2010, Salaris:2018}. The equation of state (EoS) of the highly dense core ($\sim$ $10^{6}−10^{10}\g\cm^{−3}$), is dictated by the degenerate electrons.

\noindent The precise ionic species found in the WD interior depend on the WD mass and ultimately on the WD progenitor mass. WDs of mass below $\sim 0.5\Msun$ are expected to be made of helium and carbon~\cite{Fontaine:2001}. WDs with masses in the range $0.5\Msun\lesssim\Mstar\lesssim 1.05\Msun$ have a carbon/oxygen degenerate core~\cite{Fontaine:2001,Salaris:2009,Salaris:2018}. WDs of mass above $\sim1.05\Msun$ are primarily made of a mixture of oxygen and neon~\cite{Siess:2007}. Aside from the aforementioned ions trace amounts of elements heavier than helium are also expected to be present in the WD core.



\subsection{Local WD sample}
\label{sec:wd_sample}

\noindent Recently, the Gaia mission has provided us with an unprecedentedly large sample of WD candidates, $\sim 359000$ high-confidence WDs~\cite{GentileFusillo:2021} (candidates with no actual spectroscopy data). Accurate parallax measurements and photometry by the second and the early third data releases of Gaia DR2~\cite{GaiaDR2:2018} and EDR3~\cite{GaiaDR3:2020}, enable the calculation of absolute magnitudes. Hence, WDs can be precisely placed in the observational Hertzsprung-Russell (HR) diagram. The observational HR diagram relates spectral type (effective temperature) to absolute magnitude (luminosity). 
The exact position of a WD in this diagram mainly depends on its mass, composition, and age. 
Spectroscopy sheds light on the composition of the WD atmosphere as well as measures the surface gravity and effective temperature $\Teff$ of the WD surface. 
Using model atmospheres for the particular WD spectral type, the mass, radius, core temperature, and cooling age for a given WD can be inferred. 

\noindent Since we are interested in detecting signals of DM annihilating into photons, we focus on WDs in the local bubble. Specifically, we consider  WDs that lie within 13 pc of the Sun, a subset of the sample of nearby WDs whose parameters were calculated in ref.~\cite{Limbach:2022} and are listed in Table~\ref{tab:source_datails}. We selected old (hence cold), isolated WDs with non-observed magnetic fields. 

\begin{figure}
    \centering
    \includegraphics[width=0.49\linewidth]{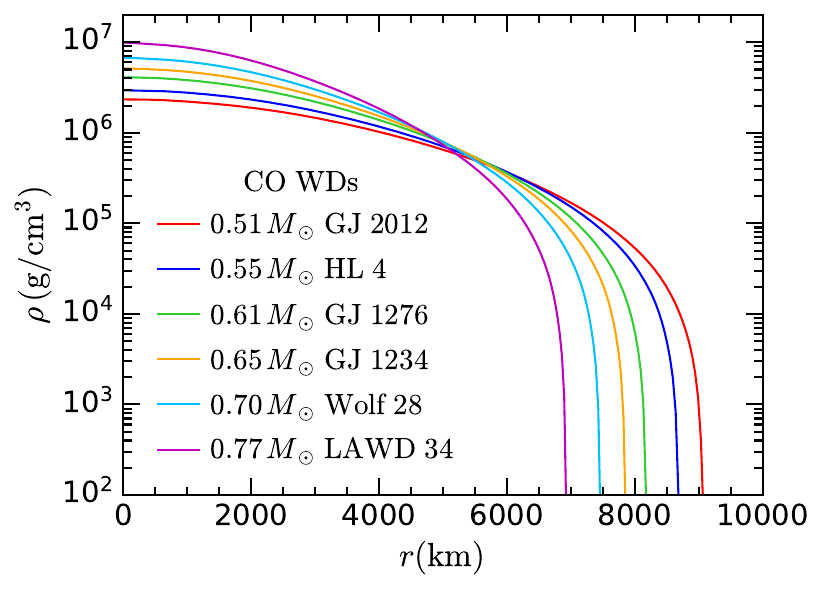}
     \includegraphics[width=0.5\linewidth]{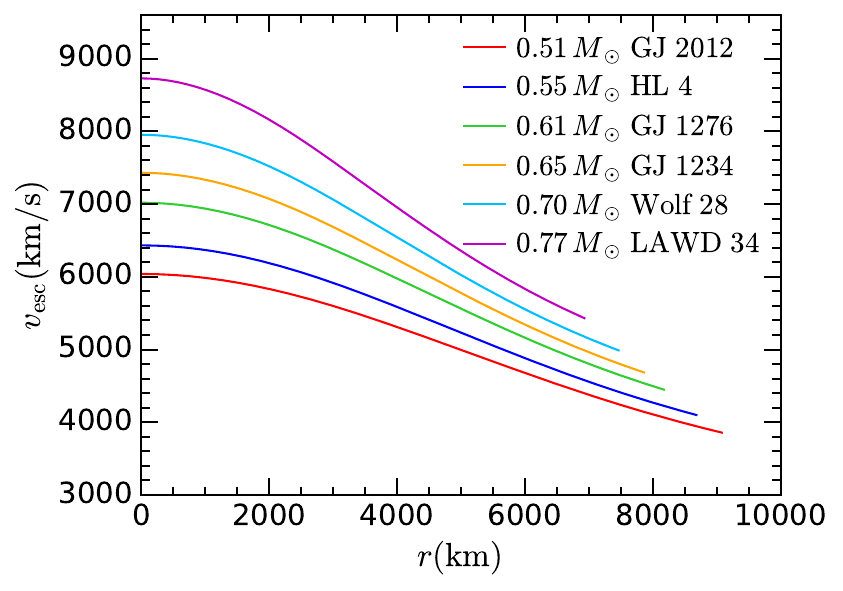}   
    \caption{Radial profiles for the core density (left) and escape velocity (right) for six of the ten sources in Table~\ref{tab:source_datails}. Profiles were calculated for the core temperatures in the second column of Table~\ref{tab:wd_evaporation}.}
    \label{fig:WDprofiles}
\end{figure}

\noindent We model the core of these WDs as a stratified mixture of equal fractions of carbon-oxygen as in ref.~\cite{Bell:2024qmj}. To this aim,  we solve the Tolman-Oppenheimer-Volkoff (TOV) equations~\cite{Tolman:1939jz, Oppenheimer:1939ne} coupled with the relativistic Feynman-Metropolis-Teller (FMT) EoS~\cite{Rotondo:2009cr,Rotondo:2011zz}. Since our local sample contains WDs of mass $<0.8\Msun$, the mass range for which the zero temperature approach for the WD EoS predicts a mass-radius relation that deviates from observations, we adopt the relativistic FMT EoS at finite temperature~\cite{deCarvalho:2013rea}. To solve the EoS for C and O aside from a central pressure or energy density, the core temperature, $\Tstar$, is another required input. We estimate $\Tstar$ for the DA (hydrogen-rich atmosphere) WDs using the cooling models for WDs made of carbon and oxygen given in ref.~\cite{Bedard:2020}\footnote{The sequences are publicly available at: \url{https://www.astro.umontreal.ca/~bergeron/CoolingModels/}.}. For the remaining WDs, we assume a core temperature ${\cal O}(10^6\K)$ according to their mass, see the second column of Table~\ref{tab:wd_evaporation}. In Fig.~\ref{fig:WDprofiles}, we show radial profiles of the density and escape velocity for six of the ten sources in Table~\ref{tab:source_datails}.

\begin{table}[tb]
\centering
   \setlength{\tabcolsep}{0.35em}   
\begin{tabular}
{|l|cccccccc|}
\hline 
WD Name (ID) & RA & DEC & 
Dist. & Mass & Radius  & $\Teff$ & Age & Spectral \\
& (deg) & (deg) &(pc) & ($M_{\odot}$) & ($R_{\oplus}$) & (K) & (Gyr) & Type \\
\hline
\hline
Gaia DR2 (J2151+5917) & 327.8750 & 59.2917 & 
8.46 & 0.57 & 1.39 & 5095 & 5 & DAH \\
(Source 1) & & & & & & & & \\
\hline
GJ 1234 (J1917+386) & 289.7443 & 38.7226 & 
11.87 & 0.65 & 1.26 & 6140 & 3.55 & DC \\
(Source 2) & & & & & & & &  \\
\hline
GJ 1276 (2251-070) & 343.4724 & -6.7818 & 
8.54 & 0.61 & 1.31 & 4170 & 8.48 & DZ \\
(Source 3) & & & & & & & &  \\
\hline
GJ 2012 (0038-226) & 10.3584 & -22.3506 & 
9.10 & 0.51 & 1.44 & 5210 & 4.44 & DQpec \\
(Source 4) & & & & & & & &  \\
\hline
GJ 3182 (0245+541) & 42.1518 & 54.3896 & 
10.87 & 0.62 & 1.32 & 4980 & 7.24 & DAZ \\
(Source 5) & & & & & & & & \\
\hline
HL 4 (0552-041) & 88.7897 & -4.1686 & 
6.44 & 0.55 & 1.39 & 4430 & 7.89 & DZ\\
(Source 6) & & & & & & & & \\
\hline
LAWD~34 (1055-072) & 164.3964 & -7.5231 & 
12.28 & 0.77 & 1.11 & 7155 & 2.93 & DC\\
(Source 7) & & & & & & & & \\
\hline
WD 0821-669 (0821-669) & 125.3613 & -67.0556 & 
10.67 & 0.53 & 1.45 & 4808 & 6.58 & DA \\
(Source 8) & & & & & & & & \\
\hline
Wolf~28 (0046+051) & 12.2912 & 5.3886 & 
4.32 & 0.70 & 1.20 & 6106 & 3.3 & DZ \\
(Source 9) & & & & & & & & \\
\hline
Wolf~489 (1334+039) & 204.1327 & 3.6794 & 
8.24 & 0.54 & 1.43 & 4971 & 5.02 & DA \\
(Source 10) & & & & & & & & \\
\hline
\end{tabular}
\caption{Source Details: Column I: Source name and ID number for this analysis;
Column II \& III: RA and DEC of our sources in degrees, respectively; 
Column IV: Distance to the local WDs in pc; Column V: Mass in $M_{\odot}$; Column VI: Radius in $R_{\oplus}$; Column VII: Effective temperature in K; Column VIII: Cooling age in Gyr; Column IX: Spectral type. We took all the parameters from ref.~\cite{Limbach:2022}.}
\label{tab:source_datails}
\end{table}

\section{Dark Matter capture in WDs}
\label{sec:capture}

\noindent We consider DM that collides with the non-relativistic, non-degenerate ionic constituents of a WD. Provided that enough energy is lost in the scattering process(es), the DM will become gravitationally bound to the WD. Thus, a DM population accumulates in the WD center over time, which eventually annihilates into Standard Model (SM) particles. In what follows, we adopt the refined approach to DM capture in WDs given in refs.~\cite{Bell:2021fye, Bell:2024qmj}.

\subsection{Capture rate}
\label{sec:ionthin}

\noindent We begin by considering the simplest scenario where a single collision is enough to capture DM. In this case, the capture rate in the optical limit is given by~\cite{Bell:2021fye}
\begin{equation}
    C_1 = \frac{\rho_\chi}{m_\chi}\int_0^{R_\star} dr 4\pi r^2
    \int_0^{\umax} du_\chi \frac{w(r)}{u_\chi} \fMB(u_\chi) \,\Omega^-(w,r),\label{eq:ioncapdef}
\end{equation}
\noindent where $\rho_\chi$ is the local DM density that we take to be $\rho_{\chi} = 0.4\GeV\cm^{−3}$, $m_\chi$ is the DM mass, $\Rstar$ is the stellar radius, $u_\chi$ is the DM velocity far from the WD,  $w^2(r)=u_\chi^2+v_{\text{esc}} (r)^2$  is the DM velocity before the collision and $\vesc$ the escape velocity, $\fMB(u_\chi)$ is the DM velocity distribution in the Galactic halo truncated at the halo escape velocity $\umax$, which we assume to be Maxwell-Boltzmann and reads~\cite{Busoni:2017mhe}
\begin{equation}
    \fMB(u_\chi)du_\chi=\frac{u_\chi}{v_dv_\star}\sqrt{\frac{3}{2\pi}}
    \left(\exp\left[-\frac{3}{2v_d^2}\left(u_\chi-v_\star\right)^2\right]-\exp\left[-\frac{3}{2v_d^2}\left(u_\chi+v_\star\right)^2\right]\right)du_\chi,  
\end{equation}
\noindent where $\vstar$ is the WD velocity in the Galactic rest frame, and $v_D$ is the velocity dispersion of the halo.  The most general expression for the scattering rate, $\Omega^-(w)$,  including thermal effects reads 
\begin{align}
    \Omega^-(w,r) &= \int_0^{\vesc(r)}dv R^-(w(r)\to v),\label{eq:iongammadef}\\
    R^-(w\to v) &=\int_0^\infty ds\int_0^\infty dt\frac{32 \mu_+^4}{\sqrt{\pi}}\kappa^3n_T(r)\frac{d\sigma_{T\chi}}{d\cos\theta_\text{cm}}\frac{vt}{w}e^{-\kappa^2v_T^2}\Theta(t+s-w)\Theta(v-|t-s|),
    \label{eq:ionRdef}
\end{align}
\noindent where $\kappa^2=m_T/2\Tstar$ with $m_T$ the target mass, $n_T(r)$ is the number density of the target nucleus at a distance $r$ from the center of the star. All kinematic quantities are defined in the center of mass (cm) frame, $s$ and $t$ are the cm velocity and the DM speed in the cm frame, $v_T$ is the target speed before the collision
\begin{equation}
    v_T^2  = 2 \mu \mu_+ t^2 + 2 \mu_+ s^2 - \mu w^2,\label{eq:cmvel}
\end{equation}
and
\begin{equation}
    \mu = \frac{m_\chi}{m_T},\qquad 
    \mu_\pm = \frac{\mu\pm 1}{2}.     
\end{equation}
\noindent It is worth noting that in the $T_\star\rightarrow0$ approximation, the scattering rate reduces to 
\begin{equation}
    \Omega^-(w,r) = \frac{4\mu_+^2}{\mu w}n_T(r)\int_{w(r)\frac{|\mu_-|}{\mu_+}}^{v_{\text{esc}(r)}} v  \frac{d\sigma_{T\chi}}{d\cos\theta_\text{cm}}(w,q_\text{tr}^2) \, dv,
    \label{eq:OmegaT0}
\end{equation}
\noindent where the dependence on the transferred momentum $q_\text{tr}$ comes from the nuclear response functions. 
For the  DM-ion differential cross section, we consider a constant DM-nucleon cross section, $\sigma_{N\chi}$, and the nuclear form factors $F(q_\text{tr}^2)$ for carbon and oxygen given in ref.~\cite{Catena:2015uha}. 
For this particular scenario Eq.~\ref{eq:OmegaT0} is well approximated by
\begin{equation}
      \Omega^-(w,r)  \simeq 2 n_T(r) \sigma_{N\chi}  \frac{(m_\chi+m_N)^2 }{ w(r) \, m_N^2} \int_{w(r)\frac{|\mu_-|}{\mu_+}}^{v_{\text{esc}(r)}} v  F(q_\text{tr}^2) \, dv,
\end{equation}
where $m_N$ is the nucleon mass.

\begin{figure}
    \centering
    \includegraphics[width=0.5\linewidth]{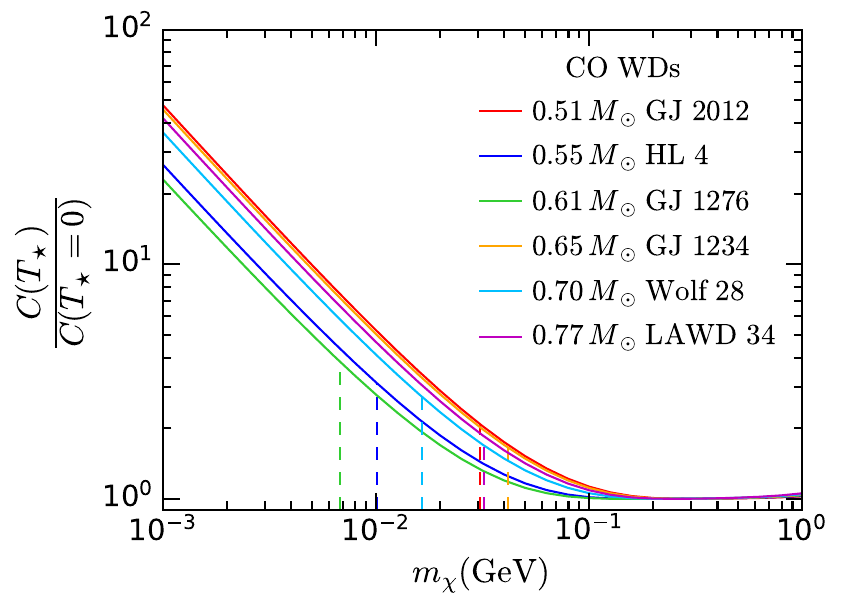}
    \caption{Finite temperature effects on the DM capture rate for six WDs in our local sample. The dashed lines denote the evaporation mass.}
    \label{fig:Cfinitetemp}
\end{figure}

\noindent The zero temperature approximation is valid when the target has a negligible velocity with respect to that of the DM. For the WDs in our local sample, this is a good approximation for $m_\chi\gtrsim0.1\GeV$ (see Fig.~\ref{fig:Cfinitetemp}), below this threshold, finite temperature effects are relevant. To consider this, we calculate the capture rate using the interaction rate in Eqs.~\ref{eq:iongammadef} and \ref{eq:ionRdef},  and make the following replacement $v_d\rightarrow\sqrt{v_d^2+3\Tstar/m_T}$ in the DM speed distribution to account for the thermal motion of the targets. 

\noindent One may think that for sub-GeV DM, effects of the Coulomb lattice structure should be accounted for in the expression of the capture rate~\cite{Acevedo:2023xnu}, specially for very light DM.  We checked that even if the WD core had completely crystallized, which is not the case of the WDs in our local sample, the lattice structure would have had no suppressing effect on the capture rate, for the mass range considered in this work. This is done by introducing the dynamic structure function of the lattice $S(q_\text{tr}) \approx 1 - e^{-2 W(q_\mathrm{tr})}$ in the differential DM-ion cross section Eq.~\ref{eq:ionRdef}, as outlined in ref.~\cite{Bell:2024qmj}. This function is defined in terms of the Debye-Waller factor $W(q_\mathrm{tr})$  given in ref.~\cite{Baiko:1995}, and is relevant only for very small momentum transfers, not realized in the DM mass range explored here.

\noindent If the DM mass is large enough, larger than ${\cal O}(\TeV)$ for local WDs,  multiple collisions with the ions in the WD core are required for the DM velocity to fall below the escape velocity of the star. In such a case, we use the following expression to compute the capture rate~\cite{Bell:2024qmj}
\begin{equation}
   C = \sum_N C_N = \frac{\rho_\chi}{m_\chi}\int_0^{R_\star} dr 4\pi r^2 n_T(r) \vesc^2(r) \sigma_{T\chi}(r)  \int_0^{\umax} du_\chi \frac{\fMB(u_\chi) }{u_\chi} \tilde{G}\left(r,\frac{m_\chi u_\chi^2}{2E_0}\right), 
   \label{eq:Ctot}
\end{equation}
\noindent where $E_0$ depends on the specific ion target and $\tilde{G}$ is the response function for multiple scattering, both quantities are defined in ref.~\cite{Bell:2024qmj}, and the DM-ion cross section is given by
\begin{eqnarray}
\sigma_{T\chi}(r) 
    &=& \frac{2}{E_R^\mathrm{max}(r)}  \int_{0}^{E_R^\mathrm{max}(r)}dE_R \frac{d\sigma_{T\chi}}{d\cos\theta_{cm}}(\vesc,E_R) \\
    &\simeq& \frac{4 \sigma_{N\chi}}{E_R^\mathrm{max}(r)}  \frac{(m_\chi+m_N^2)}{(m_\chi+m_T)^2}\frac{m_T^2}{m_N^2}\int_{0}^{E_R^\mathrm{max}(r)}dE_R F(E_R), \\
     E_R^\mathrm{max}(r) &=& 2m_T \vesc(r)^2 \left(\frac{m_\chi}{m_T+m_\chi}\right)^2, \label{eq:ER_max}   
\end{eqnarray}
\noindent where $E_R=q_\text{tr}^2/2m_T$ is the recoil energy of the target. 

\begin{figure}
    \centering
    \includegraphics[width=0.6\linewidth]{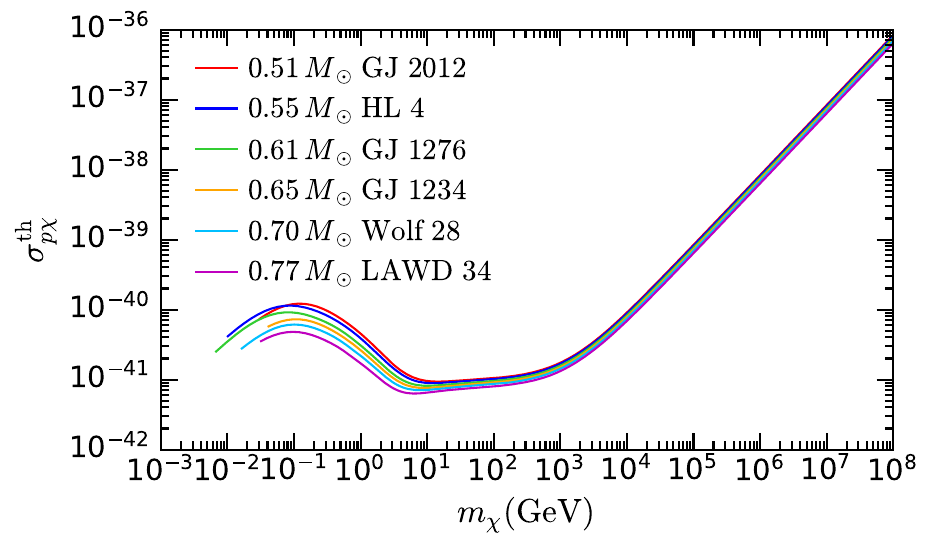}
    \caption{Geometric (threshold) DM-proton cross section, above which the DM capture rate saturates the geometric limit, for the same WDs as in Fig.~\ref{fig:Cfinitetemp}.}
    \label{fig:sigmath}
\end{figure}

\noindent There is a natural upper bound on the rate at which DM can be captured in any astrophysical object, also known as the geometric limit, which is realized when all DM particles traversing the WD are captured,
\begin{equation}
    C_\text{geom}=\frac{\pi R_\star^2 \rho_\chi}{m_\chi} \int_0^{\umax} \frac{w^2(R_{\star})}{u_\chi} \fMB(u_\chi)du_\chi.\label{eq:geom_Tto0}
\end{equation}
\noindent Thus, there is the maximum value of the DM-nucleon cross section above which the capture rate saturates the geometric limit, i.e., the capture rate becomes independent of this cross section and is equal to the value in the geometric limit. The value of this threshold (geometric) cross section can be obtained by equating either Eq.~\ref{eq:ioncapdef} or Eq.~\ref{eq:Ctot} to the expression for the geometric limit in Eq.~\ref{eq:geom_Tto0}. In Fig.~\ref{fig:sigmath}, we show this cross section for six WDs in our local sample. Note here that the DM-proton and DM-neutron cross sections are not exactly the same due to different form factors. Hereafter, our results will be given in terms of the DM-proton cross section $\sigma_{p\chi}$.  Note that the threshold cross sections have been truncated in the sub-GeV regime at the value of the evaporation mass (see text below). 

 \noindent Captured DM continues to scatter with the ions in the WD interior. If the DM mass is sufficiently low, it can gain energy in the collision and escape the WD, reducing the number of accreted DM particles. This is referred to in the literature as evaporation. We follow the prescription given in ref.~\cite{Bell:2021fye} to compute the evaporation rate and to estimate the DM mass below which this process is relevant, the so-called evaporation mass. Our results for the 10 sources are listed in Table~\ref{tab:wd_evaporation} (see also Fig.~\ref{fig:Cfinitetemp}). We also show the core temperatures used for this computation. For the DA WDs, these have been derived using the cooling models of ref.~\cite{Bedard:2020}, while for other spectral types, we assume an expected core temperature based on the WD mass and age. Specifically, based on the effective temperatures of the DC WDs in our sample ($T_\text{eff}>5000$~K), we can take them as possessing hydrogen rich atmospheres, and the DZs with $T_\text{eff}<5000$~K as helium rich~\cite{Blouin:2019}. These are conservative assumptions since DC (DZ) WDs are expected to be cooler than DA (DB), and a lower core temperature implies a lower evaporation mass. 
 In addition, we assume a DM-nucleon cross section close to the threshold value. It is worth remarking that the evaporation mass relies heavily on the core temperature and weakly on the DM-nucleon cross section~\cite{Bell:2021fye}.

\begin{table}[tb]
\centering
\begin{tabular}{|l|c|c|}
\hline 
Source & $\Tstar$ (K) & $m_{\rm evap}$~(GeV) \\
\hline
 1 (Gaia DR2 J2151+5917) & $2.7\times10^6$  & 0.0288\\
 2 (GJ 1234) & $3.4\times10^6$ &  0.0415\\
 3 (GJ 1276) & $7.6\times10^5$ &  0.0068\\
 4 (GJ 2012) & $2.5\times10^6$ & 0.0306 \\ 
 5 (GJ 3182) & $1.9\times10^6$  & 0.0168\\
 6 (HL 4) & $8.6\times10^5$ & 0.0101 \\
 7 (LAWD 34) & $4.0\times10^6$ &  0.0320 \\
 8 (WD 0821-669) & $1.8\times10^6$ &0.0208 \\
 9 (Wolf 28) & $2.5\times10^6$ & 0.0165 \\
 10 (Wolf 489) & $2.5\times10^6$ & 0.0284 \\

\hline
\end{tabular}
\caption{Evaporation mass ($m_{\rm evap}$)  of all ten local WDs and the core temperature $\Tstar$ and the core temperatures used to estimate them. 
\label{tab:wd_evaporation}} 
\end{table}

\subsection{Capture-Annihilation Equilibrium}
\label{sec:dm_annihilation}

\noindent Upon being trapped inside a WD, DM will continue to scatter and eventually settle down in a region very close to the WD center, where the thermalized DM will annihilate if self-annihilation is permitted. 
For DM masses above the evaporation mass, the number of DM particles accumulated inside the WD at a given time 
follows the equation
\begin{equation}
\frac{{\rm d}N_\chi(t)}{{\rm d}t} = C - C_{\rm ann}N_\chi^2(t). \\
\label{eqn:N_variation}
\end{equation}
\noindent Here, $C$ represents the total capture rate discussed earlier, and $C_{\rm ann}$ is related to the annihilation rate. Eq.~\ref{eqn:N_variation} can be solved analytically 
\begin{equation}
N_\chi(t) = C t_{\rm eq} \tanh\left(\frac{t}{t_{\rm eq}}\right), \label{eqn:Neq_time}
\end{equation}
\noindent where $t_{\rm eq}$, defined as $t_{\rm eq} \equiv 1/\sqrt{C_{\rm ann}C}$, represents the capture-annihilation equilibrium timescale.

\noindent The annihilation rate is defined as:
\begin{equation}
\Gamma_{\rm ann} = \frac12 C_{\rm ann} N_\chi^2. 
\end{equation}
\noindent $C_{\rm ann}$ depends on the thermally averaged annihilation $\langle\sigma_{\rm ann}v_{\chi\chi}\rangle$ and can be calculated using the general expression in ref.~\cite{Garani:2017jcj},  the parametrization $\langle\sigma_{\rm ann}v_{\chi\chi}\rangle=a+bv_{\chi\chi}^2$ and the DM number density within the WD
\begin{equation}
   n_\chi(r)  \simeq \frac{N_\chi}{\pi^{3/2} r_\chi^3} \exp{[-r^2/r_\chi^2]},   
   \label{eq:niso}
\end{equation}
\noindent where 
$r_{ \chi}$ is the radius of the isothermal sphere of DM particles in the WD core
\begin{equation}
r_{\chi}  = \sqrt{\frac{3 \Tstar}{2\pi G m_\chi \rho_c}},  
\end{equation}
\noindent and $\rho_c$ is the WD central density. 
This results in 
\begin{equation}
C_{\rm ann} = \frac{1}{(2\pi)^{3/2} r_{\chi}^3} \left[ a + \frac{2b(3\pi-8) \Tstar}{\pi \, m_\chi} \right] \, ,
\label{eqn:Cann}
\end{equation}
which contains both terms, s-wave (a) and p-wave (b).

\noindent If equilibrium  between capture and annihilation processes is reached today, i.e., $t_\star \gg t_{\rm eq}$, the total annihilation rate is solely determined by the capture rate
\begin{equation}
\Gamma_{\rm ann} = \frac{C}{2}\, .
\label{eqn:cap_ann}
\end{equation}

\noindent We have checked that thermalization and capture-annihilation equilibrium are both achieved within the lifespan of the ten WD sources (with cooling ages above $3\Gyr$)~\cite{Bell:2021fye, Bell:2024qmj}. E.g. even if the WD core is already crystallized, thermalization for DM of mass $1\TeV$ and scattering cross section $\sigma_{p\chi}=10^{-41}\cm^2$ reaches thermal equilibrium in a timescale much shorter than a year~\cite{Bell:2024qmj}. Either if we assume p-wave annihilation or s-wave using Eq.~\ref{eqn:Cann}, we find that the DM capture-annihilation equilibrium condition is quickly fulfilled, with the maximum timescale being ${\cal O}(10^4\yr)$ for p-wave annihilation and DM mass ${\cal O}(100\GeV)$.

\subsection{Spectrum}
\label{sec:dm_spectrum}

\noindent In this work, we investigate the emission of gamma rays resulting from the annihilation of DM into long-lived mediators. Typically, direct annihilation of DM into SM particles, such as quark–antiquark pairs, produces a diffuse gamma-ray spectrum through subsequent hadronization and decay processes. In scenarios where DM annihilates directly into photons or promptly decaying SM states within the WD, these products are likely to be absorbed, contributing to internal heating rather than an observable external signal.

\noindent However, the annihilation products can either contribute to heating by depositing their kinetic energy or yield detectable gamma rays and such fate depends on several parameters: the DM mass, its interaction cross section with nucleons, and the properties of the mediator, particularly its lifetime and boost. Heating typically dominates when the mediator is short-lived or the DM mass is high enough to produce energetic SM final states confined within the WD. Conversely, in the regime where DM annihilates into long-lived mediators ($\phi$) with decay lengths larger than the stellar radius, the decay products can escape the WD and yield observable gamma rays. This parameter space, where the mediator escapes before decaying, has been explored in several studies \cite{Pospelov:2007mp, Pospelov:2008jd, Batell:2009zp, Dedes:2009bk, Fortes:2015qka, Okawa:2016wrr, Yamamoto:2017ypv, Holdom:1985ag, Holdom:1986eq, Chen:2009ab, Rothstein:2009pm, Berlin:2016gtr, Cirelli:2016rnw, Cirelli:2018iax, Mazziotta:2020foa, HAWC:2018szf, Niblaeus:2019gjk, Albert:2018jwh, Andrade:2024ekx, ANTARES:2016obx, Niblaeus:2019gjk}, and forms the focus of our analysis. In this regime, the decay of mediators outside the star enhances the chances of detection by gamma-ray telescopes.

\noindent Nonetheless, there remains a possibility that these mediators might interact with SM constituents within the WD before escaping, potentially resulting in a significant decrease in the observed flux of gamma rays. As illustrated in various refs.~\cite{Leane:2021ihh, Bhattacharjee:2022lts, Bhattacharjee:2023qfi, Bhattacharjee:2024pis}, it is possible to substantially mitigate this attenuation by carefully selecting appropriate model parameters. Hence, for the sake of simplicity and ease of comparison with existing literature, we focus on the scenario where the long-lived mediator facilitates DM annihilation outside the WD, such as $\chi \chi \rightarrow \phi \phi \rightarrow 4 \gamma$. Additionally, we assume that \(m_\phi\ll m_\chi\), under which the provided formulas remain applicable even though this approximation removes the dependence on the additional parameter $m_\phi$. 
Such setup requires the mediator to decay outside the WD yet close enough that the resulting emission signal appears point-like to the detectors, which we discuss in the next section. This assumption translates into the condition $R_\star \lesssim L \lesssim d_\star \theta_{\rm res}$, where $L$ is the decay length, $R_\star$ is the WD radius, $d_\star$ the distance to Earth, and $\theta_{\rm res}$ the angular resolution of the instrument. We also stress that this range is not unnaturally narrow when translated to the model parameters. The decay length $L$ is linked with the coupling with photons $g_{\phi}$ and boost factor $\eta~=m_\chi/m_\phi$, via the lifetime $\tau$,
as $\sim  (m_\chi/m_\phi) \, c \, \tau(m_\phi, g_{\phi})$, so several orders of magnitude in $L$ are naturally spanned by modest variations in the coupling and mediator mass \cite{Leane:2017vag, Bhattacharjee:2022lts}. Moreover, this decay length window is not arbitrary but is directly motivated by observational constraints \cite{Leane:2017vag, Leane:2021ihh, Bhattacharjee:2022lts, Acevedo:2023xnu, Leane:2024bvh}. For simplicity and consistency with existing literature,  our study remains as model-independent as  possible, and we assume a 100\% branching ratio, though our considered $L$ range can be translated to well-motivated models such as the dark photon~\cite{Kyselov:2024dmi} or Higgs-portal scalar~\cite{Bernal:2025qkj}.

\noindent The differential flux of gamma rays reaching Earth as a result of DM annihilation via a long-lived mediator is given by 
\begin{equation}
\frac{d \phi_{\gamma}}{d E_{\gamma}} = \frac{\Gamma_{\rm ann}}{4 \, \pi \, d_{\star}^{2}}  \left(\frac{d N_{\gamma}}{d E_{\gamma}} \right)  P_{\rm surv},
\label{eqn:dm_flux_gamma}
\end{equation}
\noindent where $d_{\star}$ denotes the distance between Earth and the WD selected in section~\ref{sec:wd_sample} (Table~\ref{tab:source_datails}), and $P_{\rm surv}$ is the survival probability for gamma rays reaching the detectors, which is nearly unity within the range of mediator decay lengths.

\noindent Under our stated hypothesis, where $m_{\phi}\ll m_\chi$, the reliance on $m_{\phi}$ is removed, leading to the conceptualization of the gamma-ray spectrum as a box-shaped distribution, as described by ref.~\cite{Ibarra:2012dw} in Eq. \ref{eq:box_function}, i.e. 
\begin{equation}
\frac{{\rm d}N_{\gamma}}{{\rm d}E_{\gamma}} = \frac{4\Theta(E-E_{-}) \Theta(E_{+}-E)}{\Delta E}\,,
\label{eq:box_function}
\end{equation}
where $\Theta$ is the Heaviside-theta function. The energy bounds of the box are denoted by $E_{\pm}=(m_{\chi}\pm \sqrt{m_{\chi}^2-m_{\phi}^2})/2$, with the width of the box function defined as $\Delta E=\sqrt{m_{\chi}^2-m_{\phi}^2}$.

\noindent Box-shaped spectra is widely adopted in the literature when considering DM annihilation into long-lived mediators~\cite{Batell:2009zp, Ibarra:2012dw, Leane:2021tjj}. While Eq.~\ref{eq:box_function} is written for $m_\phi \ll m_\chi$, however, any mediator mass leading to a decay length $R_\star \lesssim L \lesssim d_\star \theta_{\rm res}$ will also produce $P_{\rm surv}\simeq1$  will provide essentially identical bounds.

\section{Looking for gamma-ray emission from local WDs}
\label{sec:gamma_intro}

In this section, we study the expected gamma-ray signals from WDs due to the decay of long-lived mediators produced by DM annihilation, using current and future-generation telescopes. By focusing on the DM-nucleon scattering cross section in equilibrium, we assess whether WDs could serve as viable targets for detecting gamma-ray emission linked to DM interactions. This analysis lays the groundwork for evaluating the capabilities of both existing and upcoming gamma-ray observatories in the next section. 

\noindent It is worth remarking here that for isolated white dwarfs to emit gamma rays without  DM annihilation, they need to be fast rotating and highly magnetized, hence young (for a brief review of possible production mechanisms see e.g. refs.~\cite{Meintjes:2023, Madzime:2024ulm}). Note, however, that the evidence of recent findings using \textit{Fermi} Large Area Telescope (\textit{Fermi}-LAT) for either isolated WDs or WDs in binary systems is not fully demonstrated yet~\cite{Meintjes:2023, Madzime:2024ulm}, and it would affect WDs that are not included in our sample (young highly magnetized or in binary systems).   Regarding the possible presence of unresolved wide binary systems in our sample, the expected number of such systems within 13 pc from the Sun (maximum distance of the WDs in our sample) is at most 1~\cite{Rebassa-Mansergas:2021}. 
 These systems are composed of a main-sequence star and a WD, with a separation between the companions of $\gtrsim10$~AU. Even if we had one of these systems in our sample, the separation between the companions prevents any mass transfer events (that could produce radiation emission) from happening and both components evolve as isolated stars~\cite{Willems:2004iy, Farihi:2010}. Thus, unresolved wide binaries do not affect our results.  

\subsection{Current and future gamma-ray telescopes}
\label{sec:gamma_tele}

The \textit{Fermi}-LAT in operation since 2008, has played a key role in surveying the high-energy gamma-ray sky ($E > 0.1$ GeV), enabling discoveries across sixWe analyze nearly 16 years of \textit{Fermi}-LAT data (2008-08-04 to 2024-06-04), using fermipy v1.1.0 and \textit{Fermi} Science Tools v2.2.0\footnote{\url{https://fermi.gsfc.nasa.gov/ssc/data/analysis/software/}}, with Pass 8 source-class IRFs (P8R3\_SOURCE\_V3)\footnote{\url{https://fermi.gsfc.nasa.gov/ssc/data/analysis/documentation/Pass8_usage.html}}.

\noindent We select gamma-ray events in the energy range $0.5$–$500$ GeV within a $15^{\circ}$ region of interest (ROI) centered on each WD. The model includes the target source, nearby 4FGL-DR4 sources~\cite{Fermi-LAT:2022byn}\footnote{\url{https://fermi.gsfc.nasa.gov/ssc/data/access/lat/14yr_catalog/}}, and diffuse backgrounds (gll\_iem\_v07.fits and iso\_P8R3\_SOURCE\_V3\_v1.txt)\footnote{\url{https://fermi.gsfc.nasa.gov/ssc/data/access/lat/BackgroundModels.html}}. A bin-by-bin binned likelihood analysis\footnote{\url{https://fermi.gsfc.nasa.gov/ssc/data/analysis/scitools/binned_likelihood_tutorial.html}} is performed, allowing the spectral parameters of sources within $10^{\circ} \times 10^{\circ}$ and the normalizations of diffuse components to vary. Each WD is modeled as a point source with a power-law spectrum (${\rm d}N/{\rm d}E \propto E^{-\Gamma}$) and fixed spectral index $\Gamma = 2$. The Test Statistic (TS), defined as TS $= -2\ln(L_{\rm max,0}/L_{\rm max,1})$, is used to search for excess emission. No significant detection is found (TS $<25$). In the absence of a signal, we derive 95\% confidence level upper limits on the gamma-ray flux using the profile likelihood method~\cite{Rolke:2004mj}, with $-2\Delta\ln(\mathcal{L}) = 2.71$.

\noindent High energy photons are not generally expected to be emitted by old isolated WDs. A WD (point source) emitting in gamma rays could be explained by DM annihilating into long-lived mediators that decay outside the WD (see Section~\ref{sec:dm_spectrum}). 
Thus, we can conservatively translate \textit{Fermi}-LAT differential flux upper limits for point-like sources into constraints on the DM–nucleon cross by comparing them with Eq.~\ref{eqn:dm_flux_gamma}, for a given DM mass and every nearby WD in our local sample. However, with current \textit{Fermi}-LAT sensitivity and 16 years of data, none of the WDs yield meaningful constraints. For instance, for $m_\chi = 1000$ GeV, Source 9's (Wolf 28) expected limit is~$5.05\times10^{-41}~\rm{cm^{2}}$ which exceeds its geometric threshold by a factor $\sim$~6 (see Fig.~\ref{fig:sigmath}). As discussed in Section~\ref{sec:ionthin}, a meaningful experimental bound requires the DM–nucleon cross section to lie below the geometric threshold, ensuring it falls within the linear capture regime before saturation sets in. \\

\noindent Looking ahead, the Cherenkov Telescope Array (CTA), designed for Very High Energy (VHE, $E > 0.1$ TeV) observations, promises unprecedented sensitivity and precision in ground-based gamma-ray astronomy \cite{CTAObservatory:2022mvt, Gueta:2021vrf}. With its large effective area and wide field of view, CTA will play a pivotal role in the study of both extragalactic and Galactic gamma-ray sources. The Southern Wide-field Gamma-ray Observatory (SWGO) is a planned water Cherenkov detector array for the Southern Hemisphere, currently undergoing design and prototype development. It is intended to complement existing observatories such as Large High Altitude Air Shower Observatory (LHAASO) in the Northern Hemisphere. Covering an energy range from $\sim$30 GeV to several PeV, SWGO will extend sky coverage and enhance detection capabilities when operational \cite{Conceicao:2023tfb, LHAASO:2021zta}. LHAASO itself is already operational and is designed to study cosmic rays and gamma rays across a broad energy spectrum, contributing significantly to multi-messenger astrophysics~\cite{Cao:2010zz}.

\noindent In the following section~\ref{sec:bounds_gamma}, we explore the potential of current/future generation telescopes to constrain the DM-nucleon scattering cross section, based on their simulated differential flux sensitivity from the all-sky survey for point-like searches.

\subsection{Projected limits from gamma-ray telescopes on DM interactions}
\label{sec:bounds_gamma}

\noindent In this section, we examine the expected constraints on DM interactions based on the projected sensitivity of three gamma-ray telescopes: CTA, SWGO, and LHAASO~\cite{CTAObservatory:2022mvt, Gueta:2021vrf, Conceicao:2023tfb, LHAASO:2021zta, Cao:2010zz}. These next-generation observatories are designed to significantly enhance our ability to detect high-energy gamma-ray signals, offering potential insights into DM annihilation signatures from local WDs.

\begin{figure}[t]
\centering
\includegraphics[width=0.9\linewidth]{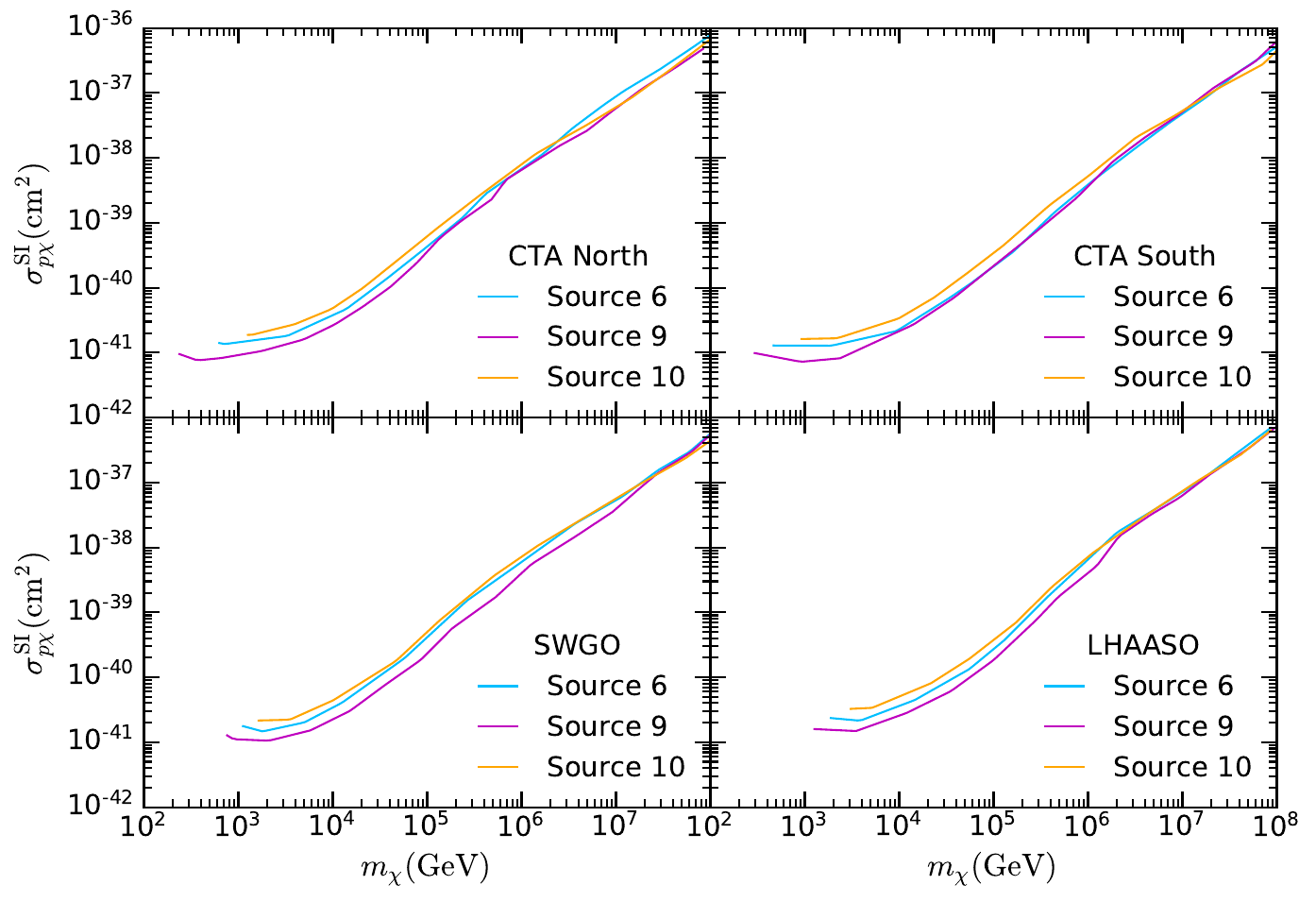}
\caption{Sensitivity projections on the DM-proton scattering cross sections from three local white dwarfs, for (a) CTA North, (b) CTA South, (c) SWGO, (d) LHAASO. } 
\label{fig:diff_flux_limits_gamma}
\end{figure}

\noindent Following the discussion of gamma-ray detectors in Section~\ref{sec:gamma_tele}, we focus on the differential flux sensitivity of these experiments towards detecting point sources. The differential sensitivity is defined as the minimum flux required to achieve a 5$\sigma$ detection of a point-like source, calculated in non-overlapping logarithmic energy bins from an all-sky survey. The differential sensitivity for CTA~\cite{CTA:2020hii, cta, CTAConsortium:2023ybn}, LHAASO~\cite{LHAASO:2019qtb, LHAASO:2021zta}, and SWGO~\cite{Albert:2019afb} is derived using a 5$\sigma$ detection threshold for point-like sources, ensuring statistically robust observations. Observation times considered are 50 hours for CTA\footnote{Performance data available at: \url{https://www.ctao.org/for-scientists/performance/}} and one and five years for LHAASO and SWGO, respectively, reflecting their wide-field monitoring capabilities.

\noindent We consider the DM annihilation process via long-lived mediators, as discussed in section~\ref{sec:dm_spectrum}, assuming a 100\% branching ratio into gamma rays. This provides the maximum upper limit on the DM capture rate, which can then be translated into a constraint on the DM–nucleon scattering cross section. In Fig.~\ref{fig:diff_flux_limits_gamma}, we present projected sensitivities on the DM-proton scattering cross section, $\sigma_{p\chi}$, as a function of the DM mass, $m_\chi$. Using Eq.~\ref{eqn:dm_flux_gamma}, we compute the expected flux as a function of $m_\chi$ and $\sigma_{p\chi}$, equating it with the projected sensitivity of the selected gamma-ray detectors. 
For clarity, we focus on the three most promising WD candidates, Sources 6, 9, and 10, since they yield the strongest projected constraints. 

\noindent Unlike current instruments such as \textit{Fermi}-LAT, which have limited sensitivity at higher DM masses, the future gamma-ray telescopes, including CTA South, CTA North, SWGO, and LHAASO, are expected to provide substantially stronger constraints for DM masses above $10^{3}$ GeV. Among these, CTA South and LHAASO emerge as the most promising observatories for probing local WDs, offering the most stringent bounds across a broad DM mass range of approximately $10^{3}$–$10^{8}$ GeV. These improved sensitivities will allow us to explore previously unconstrained parameter space, potentially shedding light on DM interactions with ions in WDs.

\subsection{Best Candidates: Comparison with Direct Detection Limits}
\label{sec:best_candidate}

\begin{figure}[t]
\centering

\includegraphics[width=0.6\linewidth]{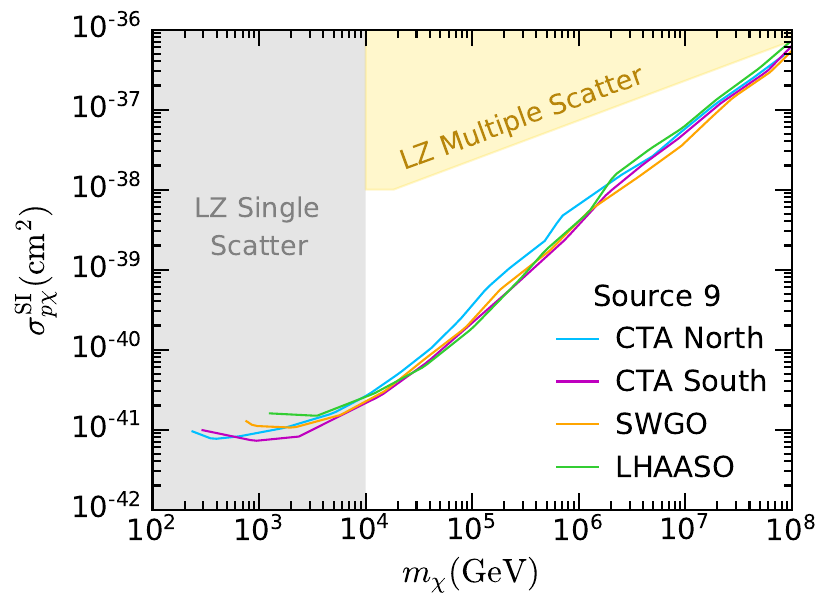}

\caption{Sensitivity projections from source 9 for future gamma-ray telescopes and comparison with the LZ spin-independent limits (shaded regions) for single~\cite{LZ:2024zvo} and multiple scattering~\cite{LZ:2024psa}.}
\label{fig:direct_comp}
\end{figure}

\noindent In Fig.~\ref{fig:direct_comp}, we compare the expected constraints from all selected gamma-ray telescopes for Source 9 (Wolf 28), which provides some of the strongest bounds among the WDs in our local sample. 

\noindent To contextualize these results, we also compare our findings with limits from leading direct detection experiments, specifically LZ single-scattering~\cite{LZ:2024zvo} and multiple-scattering~\cite{LZ:2024psa} analyses. The LUX-ZEPLIN (LZ) experiment is one of the most sensitive terrestrial searches for DM, utilizing a liquid xenon time projection chamber to detect nuclear recoil events induced by DM interactions. While single-scattering analyses target traditional weakly interacting massive particle (WIMP) searches, multiple-scattering analyses extend sensitivity to ultra-heavy DM candidates that may interact more frequently as they traverse the detector medium. By comparing our astrophysical constraints with those from LZ, we can assess the complementarity of indirect and direct detection strategies in probing DM-nucleon interactions across a wide parameter space.  

\noindent These results highlight the potential of gamma-ray observations to probe DM interactions at masses $>~10^{4}$ GeV, where direct searches lose sensitivity, although thermal DM production at such scales is limited by the unitarity limit.

\subsection{Uncertainty}
\label{sec:uncertainty}

\noindent In this section, we quantitatively assess the uncertainties that may impact our results. The sensitivity projections presented in Figs.~\ref{fig:diff_flux_limits_gamma} and \ref{fig:direct_comp} arise from the interplay between the DM capture rate, DM density profile, and DM mass range, providing bounds on the DM-nucleon cross sections from nearby local WDs. It is important to emphasize that while the choice of DM profile affects the results, the impact is relatively modest. The primary dependence stems from the local DM density, $\rho_\chi$, which can vary by approximately 30\% when adopting, for instance, a Burkert profile with parameters as in ref.~\cite{Calore:2022stf}. The uncertainty in $\rho_\chi$ remains the dominant theoretical uncertainty, a challenge shared by all local DM probes, including direct detection experiments. However, since the expected gamma-ray flux (and hence our bounds on the DM-nucleon cross section) scales almost linearly with $\rho_\chi$, the uncertainty in $\rho_\chi$ does not affect our main conclusions, it will only induce an overall shift in the projected sensitivities, by about 30\% of our results in Figs.~\ref{fig:diff_flux_limits_gamma} and \ref{fig:direct_comp}.

\noindent Additionally, systematic uncertainties associated with the observational instruments, such as atmospheric modeling and detector calibration, play a crucial role in determining the precision of our results. For example, while using the published sensitivity projections of the detectors, CTA typically exhibits flux uncertainties in the range of 10-20\%, with additional contributions from calibration uncertainties (5-10\%) and atmospheric effects (10-15\%) \cite{CTAConsortium:2017dvg, Bernlohr:2008kv}. Similarly, LHAASO flux uncertainties are estimated at 15-20\%, with calibration uncertainties around 10\% and atmospheric modeling effects contributing 10-15\% \cite{DiSciascio:2016rgi, Wang:2023qeg}. SWGO also faces flux uncertainties of approximately 15-20\%, with calibration and atmospheric effects each contributing 10-15\%~\cite{Albert:2019afb}. 

\noindent While these instrumental uncertainties introduce additional variation in our results, their impact remains within a manageable range compared to the dominant theoretical uncertainties. Future improvements in calibration and modeling could further refine these bounds, enhancing the robustness of our sensitivity projections on DM interactions using local WDs.

\section{Conclusion and Discussion}
\label{sec:conclusions}

\noindent In recent decades, WDs have been extensively studied, with the Gaia mission significantly expanding the catalog of nearby WD candidates, including those with temperatures below 5000~K. WDs in DM-rich environments such as the Galactic Center and dwarf galaxies are ideal probes for DM capture studies; however, they either lack sufficient WD observations or have uncertain DM content. Given these limitations, focusing on WDs in the local bubble is more advantageous. Although they may not provide the most stringent bounds, their proximity significantly reduces uncertainties, leading to more reliable and robust results.  

\noindent In this study, we looked for signals of DM annihilation into long-lived mediators in the core of old (hence cold) WDs, that later decay to gamma rays outside the WD. We selected a sample of ten old, isolated WDs within 13 pc of the Sun, with no observed magnetic fields. For these WDs, no high energy photon emission is expected from SM origin.

\noindent Given the mass range of our local WD sample ($0.5\Msun<M_\star<0.8 M_\odot$), we modeled their core as a stratified mixture of equal fractions of carbon and oxygen, solving the TOV equations coupled with the finite temperature relativistic FMT equation of state. 
The WD core temperature was estimated using cooling models for DA (hydrogen-rich atmosphere) WDs, while for other WD spectral types, we assumed a core temperature of $\mathcal{O}(10^6\text{ K})$ based on their mass, age, and spectral type.  

\noindent Using these models, we computed the DM capture rate in each of these objects, which not only determines the expected gamma ray flux, but also allows us to estimate the maximum (\textit{geometric}) cross section that can be probed with these compact objects. We also 
determined the DM mass below which evaporation significantly reduces DM accumulation in these WDs. As all the WDs in our sample have core temperatures $\mathcal{O}(10^6\text{ K})$, they are sensitive to DM of mass above $\mathcal{O}(10\MeV)$ being captured. 

\noindent Our setup focuses on the regime where capture DM annihilates into long-lived mediators that eventually decay outside the WDs yet within the angular resolution of gamma-ray telescopes, ensuring a detectable and localized signal. Although our framework deliberately intends to be as model-independent as possible, and considered only light ($m_{\phi}~\ll~m_{\chi}$) and long-lived mediators, the formulae
reported are also valid for modest variations in the coupling and mediator mass. Moreover, this decay length window is not arbitrary but is directly motivated by observational constraints~\cite{ Leane:2017vag, Leane:2021ihh, Bhattacharjee:2022lts, Acevedo:2023xnu, Leane:2024bvh} and also naturally allowed in several well-motivated beyond-Standard Model scenarios, including dark photons~\cite{Kyselov:2024dmi}, Higgs portal scalars~\cite{Bernal:2025qkj}, and extended SUSY models~\cite{Pospelov:2008jk}. These setups can easily accommodate the required mass–coupling combinations, and the parameter space remains compatible with the model-dependent limits driven from cosmological constraints such as those from Big Bang Nucleosynthesis~\cite{Pospelov:2010hj} and the Cosmic Microwave Background~\cite{Slatyer:2015jla}. 

\noindent We then search for high-energy gamma-ray emission from nearby WDs using 16 years of \textit{Fermi}-LAT data. We also examine the sensitivity of current and future observatories—LHAASO, CTA, and SWGO—which offer improved reach at very and ultra-high energies. 
The null detection of point sources allows us to set upper limits on the flux expected from captured DM. From the expected gamma-ray emission via DM annihilation to long-lived mediators, we derive constraints on the DM–nucleon cross section.  Our main results are:
\begin{itemize} 
\item With the current \textit{Fermi}-LAT sensitivity and 16 years of data, no bounds on the DM-nucleon cross section from local WDs can be derived.     

\item  However, next-generation telescopes such as CTA South, CTA North, SWGO, and LHAASO can impose stronger constraints for DM masses $m_\chi > 10^{3}$ GeV ($\sigma_{p\chi} \sim 10^{-41}~\text{cm}^2$). Among these, CTA South and LHAASO emerge as the most promising telescopes, providing the strongest sensitivity projections across a broad DM mass range ($\sim 10^{3}-10^{8}$ GeV). The most stringent result, particularly from CTA South at $m_\chi \sim 10^3$ GeV, is obtained for Wolf 28, a $0.7\Msun$ WD at $4.3$~pc from us.   

\item Our results for DM of mass larger than $10\TeV$ are competitive with leading direct detection experiments, particularly LZ multiple scattering~\cite{LZ:2024psa}, highlighting the potential of WDs as complementary targets for DM searches.  
\end{itemize}

\noindent Local WDs provide a unique and robust avenue for DM indirect detection, complementing other astrophysical and direct detection searches for DM models such as the one explored in this work. Future observational campaigns with upcoming telescopes will enable a more systematic survey of these objects. A larger sample size, combined with improved theoretical modeling, could significantly refine constraints on DM interactions, particularly in mass ranges where other methods struggle to reach sensitivity.

\noindent Our study highlights the potential of local WDs as promising astrophysical targets for DM searches. While current telescopes cannot probe cross sections smaller than the geometric value for sub-GeV DM, upcoming gamma-ray observatories are poised to refine current sensitivity projections, particularly for high-mass DM candidates. The synergy between indirect detection using WDs and direct detection experiments will be crucial in narrowing down the parameter space for DM interactions. Furthermore, our results emphasize the importance of systematic studies of WDs as complementary probes in the broader landscape of DM searches, alongside direct detection and other astrophysical methods.

\section*{Acknowledgements}
We thank Michael Virgato for providing the equations of state and TOV solver. PB acknowledges support from the COFUND action of Horizon Europe’s Marie Sklodowska-Curie Actions research programme, Grant Agreement 101081355 (SMASH). PB would like to thank Raghuveer Garani for his valuable suggestions. SR was supported by the Fermi National Accelerator Laboratory (Fermilab), a U.S. Department of Energy, Office of Science, HEP User Facility. This work was performed in part at Aspen Center for Physics, which is supported by National Science Foundation grant PHY-2210452. SR acknowledges CERN TH Department for its hospitality while this research was being carried out. SMB was supported by the Australian Research Council through Discovery Project DP220101727.

\bibliographystyle{JHEP}
\bibliography{white_dwarf_210125}
\end{document}